\documentclass[11pt,draftcls,onecolumn]{IEEEtran}
\usepackage[cmex10]{amsmath}
\usepackage{adjustbox}
\usepackage{tabularx}
\usepackage{subfig}
\usepackage{array}
\newcolumntype{L}{>{\centering\arraybackslash}m{3cm}}
\usepackage{multicol}
\usepackage{tabularx}
\usepackage{enumerate}
\usepackage{graphicx}
\usepackage{float}
\usepackage{booktabs}
\usepackage{color}
\usepackage{amssymb}
\usepackage{epstopdf}
\UseRawInputEncoding
\DeclareGraphicsExtensions{.eps,.pdf,.png}

\usepackage[switch]{lineno}
\begin{document}
\setlength{\parskip}{0pt}
\title{Explainable Machine Learning: The Importance of a System-Centric Perspective}

\author{Manish Narwaria
\thanks{M. Narwaria is with Department of Electrical Engineering, Indian Institute of Technology Jodhpur, NH 62, Surpura Bypass Rd, Karwar, Rajasthan 342030, India. e-mail: narwaria@iitj.ac.in} 
\thanks{}}

\maketitle

\section{Scope}

The landscape in the context of several signal processing applications and even education \cite{9418577} appears to be significantly affected by the emergence of machine learning (ML) and in particular deep learning (DL). The main reason for this is the ability of DL to model complex and unknown relationships between signals and the tasks of interest. Particularly, supervised DL algorithms have been fairly successful at recognizing perceptually or semantically useful signal information in different applications (eg. identifying objects or regions of interest from image/video signals or to recognize spoken words from speech signal i.e. speech recognition etc.). In all of these, the training process uses labeled data to \emph{learn} a mapping function (typically implicitly) from signals to the desired information (class label or target label). The trained DL model is then expected to correctly recognize/classify relevant information in a given test signal. A DL based framework is therefore, in general, very appealing since the features and characteristics of the required mapping are learned almost exclusively from the data without resorting to explicit model/system development. 

The focus on implicit modeling however also raises the issue of lack of explainability/interpretability of the resultant DL based mapping or the black box problem. As a result, explainable ML/DL is an active research area \cite{xie2020explainable}, \cite{BAI2021108102}, \cite{8947948}, \cite{xDNN} where the primary goal is to elaborate how the ML/DL model arrived at a prediction. We however note that despite the efforts, the commentary on black box problem appears to lack a technical discussion from the view point of: a) its origin and underlying reasons, and b) its practical implications on the design and deployment of ML/DL systems. Accordingly, a reasonable question that can be raised is as follows. Can the traditional system-centric approach (which places emphasis on explicit system modeling) provide useful insights into the nature of black box problem, and help develop more transparent ML/DL systems?

\section{Context and relevance}
The answer to the mentioned question is a yes. This can be better understood by differentiating between a system-centric approach and a data-centric paradigm \cite{9418577}. The former in general aims at explicit modeling of the physical process by relying on apriori information and a more analytical perspective. For instance, the characterization of noise as high frequency components, use of gradient or edge (high frequency) information for shape analysis in image/video signals, exploiting correlation between signal samples (say for signal compression), locating the test statistic on a known probability density function (eg. in hypothesis testing) etc. As a result, system design philosophy and performance analysis remain largely amenable to scrutiny. In contrast, the data-centric approach (i.e. ML/DL) typically focuses on implicit system modeling by learning a mapping function from input to desired output. This is particularly aided by powerful modeling capabilities of DL \cite{10.5555/3086952} that offer the flexibility of evolving a suitable mapping function i.e. determining a set of weights from the training data. However, a direct interpretation of the mapping learnt by DL is in general difficult giving rise to the black box problem. Hence, it is reasonable to ask the stated question in the context of how a system-centric approach can help in better understanding of the black box problem in ML/DL and its practical implications. This is expected to be crucial for making meaningful progress toward development of more transparent and explainable ML systems.         
  
Therefore, the primary purpose of this lecture note is to shed light on the stated aspects of explainable ML. We also attempt to provide some perspectives on how to mitigate it from the view point of ML system design. To achieve these objectives, we rely on a system-centric philosophy to develop our arguments. We limit ourselves to an easy to understand yet meaningful example of a simple low pass filter. This, in our opinion, is not only convenient but also makes the lecture note accessible to readers from diverse backgrounds.        
\section{Prerequisites}

This lecture note assumes familiarity with basic concepts in Signals and Systems.

\section{Problem statement and solution}
Let $g:\mathbb{R}^p \rightarrow \mathbb{R}^q$ denote the mapping function from an input ${\boldsymbol{x}} \in \mathbb{R}^p$ to actual (desired) output ${\boldsymbol{y}} \in \mathbb{R}^q$. For instance, consider the application of object detection in images where we wish to recognize which of say three objects of interest is present in a $100\times 100$ image. In this case, we have $p=100\times 100$ and $q=1$ i.e. the output $g(\boldsymbol{x})$ is either $0$ or $1$ or $2$ corresponding to one class (object) label. Similarly, in the scenario of object localization, we wish to determine the location of an object in the image. We may denote this by a bounding box specified by a set of four coordinate points. Accordingly, $p=100\times 100$ and $q=8$. We note that the mapping function $g$ in both the stated applications, and indeed in many others, is typically unknown. This is where DL in particular has gained popularity since it can potentially \emph{learn} a mapping $\hat{g}$ from a set of labeled data. One then hopes that $\hat{g}$ is an accurate estimator of $g$ from a practical view point.

\begin{figure}[]
\center
\includegraphics[scale=0.62]{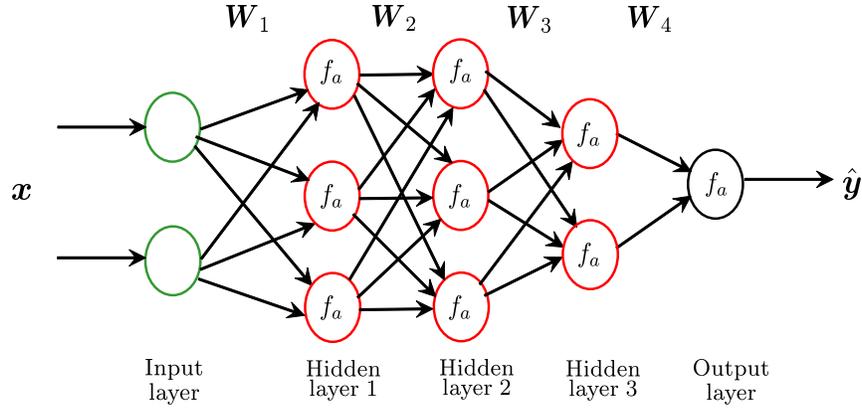}
\caption{A fully connected feed-forward network with $3$ hidden layers ($L=3$). First two hidden layers have 3 neurons while the third has 2 neurons. As $p=2$ and $q=1$ in this example, the input and output layers have 2 and 1 neurons, respectively. $f_a$ denotes the activation function and $\boldsymbol{\theta} = \left[\boldsymbol{W}_1,...,\boldsymbol{W}_{4}\right]$ is the parameter learned via training. The entries of $\boldsymbol{W}_1,...,\boldsymbol{W}_{4}$ denote the weights of the connections between the neurons. Refer to Figure \ref{DLnetworks} for an example.}
\label{DLarchitecture}
\end{figure}

At this point, it would be instructive to describe briefly the basic working of DL using notations. As mentioned, we denote the input to the DL model as ${\boldsymbol{x}} \in \mathbb{R}^p$ i.e. ${\boldsymbol{x}}$ is a vector of length $p$. Then, the $q$-dimensional output of the DL model ${\boldsymbol{\hat{y}}} \in \mathbb{R}^q$ is:
\begin{equation}
\boldsymbol{\hat{y}} = \hat{g}\left(\boldsymbol{x};\boldsymbol{\theta}\right)\,,
\end{equation}
Here, $\boldsymbol{\theta} = \left[\boldsymbol{W}_1,...,\boldsymbol{W}_{L+1}\right]$ denotes the parameter consisting of $L+1$ weight matrices (for simplicity, we ignore the intercept or bias term). The entries of matrices $\boldsymbol{W}_1,...,\boldsymbol{W}_{L+1}$ denote the weights of the connection between different neurons present in a DL model/architecture \cite{10.5555/3086952}. These neurons are arranged in a DL model through $L$ hidden layers. As an illustration, we show in Figure \ref{DLarchitecture} a DL model with 3 hidden layers i.e $L=3$. Notice that in this example the first two hidden layers have $3$ neurons while the third has $2$ neurons. Moreover, let us assume assume that $p=2$ and $q=1$. Hence, the input and output layers have $2$ and $1$ neurons, respectively. The reader can observe from Figure \ref{DLarchitecture} that there are 4 weight matrices: $\boldsymbol{W}_1$, $\boldsymbol{W}_2$ and $\boldsymbol{W}_3$ correspond to the three hidden layers while $\boldsymbol{W}_4$ represents the weight matrix for the output layer. To compute the output $\boldsymbol{\hat{y}}$, the DL model uses $\boldsymbol{W}_1,...,\boldsymbol{W}_4$ in series. Also, a non linearity is introduced at each neuron via the use of the function $f_a:\mathbb{R}^c \rightarrow \mathbb{R}^c$. It is the element-wise nonlinear function commonly referred to as an activation function. Thus, the output can be written as:  
\begin{equation}
\boldsymbol{\hat{y}}  = f_a\Bigg(\boldsymbol{W}_4f_a\bigg(\boldsymbol{W}_3f_a\Big(\boldsymbol{W}_2f_a\big(\boldsymbol{W}_1\boldsymbol{x}\big)\Big)\bigg)\Bigg)\,,
\end{equation}
Note that because we have chosen $p=2$ and $q=1$ for the example in Figure \ref{DLarchitecture}, the dimensions of $\boldsymbol{x}$, $\boldsymbol{W}_1$, $\boldsymbol{W}_2$, $\boldsymbol{W}_3$, $\boldsymbol{W}_3$ and $\boldsymbol{\hat{y}}$ will be $1\times 2$, $2 \times 3$, $3\times3$, $3\times2$, $2\times1$ and $1\times1$, respectively. The goal of training the DL model is \cite{10.5555/3086952} to find the parameter $\boldsymbol{\theta} = \left[\boldsymbol{W}_1,...,\boldsymbol{W}_{L+1}\right]$ via training on a set of labeled data such that $\boldsymbol{\hat{y}}$ is close to $\boldsymbol{{y}}$ (the actual or desired output). 
\subsection{Problem statement}

We note that there are two aspects of a DL system namely design and validation. The former refers to the choice of DL architecture (eg. activation function $f_a$, number of layers $L$, number of neurons in each layer etc.) and subsequent optimization \cite{10.5555/3086952} to find the parameter $\boldsymbol{\theta}$. The latter refers to application-specific benchmarking of the trained DL model on independent test set. Thus, both the aspects of DL model depend heavily on data. As already mentioned, we refer to this as a data-centric approach. Consequently, one typically relies only on implicit modeling (i.e. without the need for explicit signal analysis or handcrafted signal features) and the prediction accuracy as surrogates to explicit system analysis. Such data-centric approach is in contrast to a system-centric approach which is typically based on explicit modeling and apriori knowledge. Therefore, our problem statement follows naturally and can be stated as: what additional and practically useful insights can a system-centric approach reveal which can eventually help in the design of more transparent and explainable ML/DL systems?  
\subsection{Solution}
To identify practically meaningful insights about explainability aspects of a DL system, we rely on the idea of exploiting explicit apriori knowledge which is fundamental to the system-centric approach. For instance, channel modeling in communication systems can exploit knowledge of probabilistic model for the channel filter taps (eg. using Rayleigh distribution). Similarly characterizing visual signals at different frequencies and orientations (eg. using Gabor filters) attempts to explicitly mimic frequency and orientation selectivity of the human visual system in different applications (eg. texture analysis). Hence, the system-centric approach typically attempts to approximate $g$ via explicit characterization of different subsystems (components) of the physical process/application under consideration. This in turn allows a multi-dimensional analysis of strengths and weaknesses of each subsystem explicitly from the viewpoint of certain established knowledge base in the context of the application. Inspired by this philosophy, we pursue the idea of using an explicit system $S$ (with known $g$) as a reference. We then attempt to model $S$ using DL i.e. train a DL network such that the learned function $\hat{g}\approx g$. Because $S$ is explicit by choice, an analysis of the resultant DL based model of $S$ can reveal additional insights in the context of our problem statement. 

Several choices of $S$ are possible. But in this lecture note we select a simple filtering application where we wish to attenuate perceptually less relevant signal information. This is a common use-case in both traditional (eg. signal denoising, smoothing, anti-aliasing etc.) and recent application areas (eg. spatial audio rendering in Augmented and Virtual Reality, design of smart cameras for IoT, medical imaging and so on).  Accordingly, we let $S$ to be a low pass (moving average) filter, and use it, as an example, to filter out signal information beyond $2\: \text{kHz}$. Apart from its conceptual simplicity, the said choice of $S$ also enables a fairly straightforward DL based implementation from the perspective of generating training data and subsequent optimization. Now, from a system-centric perspective, $S$ is conveniently described by the following difference equation:
\begin{equation}
{\boldsymbol{y}}[n] = \frac{1}{M}\sum_{k=0}^{M-1}{\boldsymbol{x}}[n-k] = \sum_{k=0}^{M-1}{\boldsymbol{w}}[k]{\boldsymbol{x}}[n-k]\,,
\label{firlpf}
\end{equation}  
where $k=0,1,...,M-1$. Note that the system $S$ denoted by \eqref{firlpf} is explicit and can also be visualized in the frequency domain for clearer physical interpretation. Setting $M=2$, we arrive at a simple low pass filter with filter coefficients ${\boldsymbol{w}}[0]={\boldsymbol{w}}[1]=0.5$. Therefore, the output is simply the average of the present and past sample in the input $\boldsymbol{x}[n]$ i.e.
\begin{equation}
\boldsymbol{y}[n]=0.5\boldsymbol{x}[n]+0.5\boldsymbol{x}[n-1]\,,
\label{firlpfndomain}
\end{equation}

We now attempt to model $S$ using DL based regression. To that end, we denote the training data as $\left\{\left({\bf{z}}_i,{\bf{t}}_i\right)\right\}_{i=1,...,T}$. Because $M=2$, ${\bf{z}}$ is a $T \times 2$ matrix with ${\bf{z}}_i$ being the $i^{th}$ row. ${\bf{t}}$ denotes an $T\times 1$ vector of the target (desired) values. Given our problem setting, ${\bf{t}}_i$ simply denotes the average of the two numbers in ${\bf{z}}_i$. $T$ represents the size of the training data. The training process seeks to estimate the function $\hat{g}$ such that ${\bf{t}}_i = \hat{g}\left({\bf{z}}_i,{\boldsymbol{\theta}}^*\right)$ where the parameter ${\boldsymbol{\theta}}^*$ minimizes a chosen loss function $\ell$ i.e.
\begin{equation}
{\boldsymbol{\theta}}^* = \arg\min_{\boldsymbol{\theta}}\ell\left({\bf{z}},{\bf{t}}, {\boldsymbol{\theta}}\right)\,,
\end{equation}
We employed the widely used MSE as the loss function i.e.
\begin{equation}
\ell\left({\bf{z}},{\bf{t}},{\boldsymbol{\theta}}\right) = \frac{1}{T}\sum_{i=1}^{T}||{\bf{t}}_i-\hat{g}\left({\bf{z}}_i,{\boldsymbol{\theta}}\right)||^2\,,
\label{loss_mse}
\end{equation}
Accordingly, we seek a mapping function such that the sum of squared difference between the predicted value $\hat{g}\left({\bf{z}}_i,{\boldsymbol{\theta}}\right)$ and the corresponding target value ${\bf{t}}_i$ is minimized i.e. $\ell\left({\bf{z}},{\bf{t}},{\boldsymbol{\theta}}\right)\leq \epsilon$. Here, $\epsilon$ represents the tolerance or the maximum error allowed during model training. Setting an appropriate value of $\epsilon$ is crucial and typically depends on applications as well as the size of training data $T$. In this lecture note, we are not overtly concerned about these aspects, and simply choose $\epsilon = 10^{-4}$ and $T=1000$ for our experiments. In addition, a complete specification of DL based model/architecture requires several hyper parameters \cite{10.5555/3086952} including the number of hidden layers $L$, number of neurons in each hidden layer, the type of activation function $f_a$, and the loss function $\ell$. Hence, these need to be chosen before a DL model can be trained. Once the DL model is trained properly, one expects that it will generalize well to data that did not appear in the training set. In other words, it is hoped that the prediction $\hat{g}\left({\bf{z}}_{test},{\boldsymbol{\theta}}\right)$ for any test signal ${\bf{z}}_{test}$ is close to the unknown target $y_{test}$ (i.e. similar to how the trained DL model behaved for the training data). 
\begin{figure}[]
\center
\includegraphics[scale=0.73]{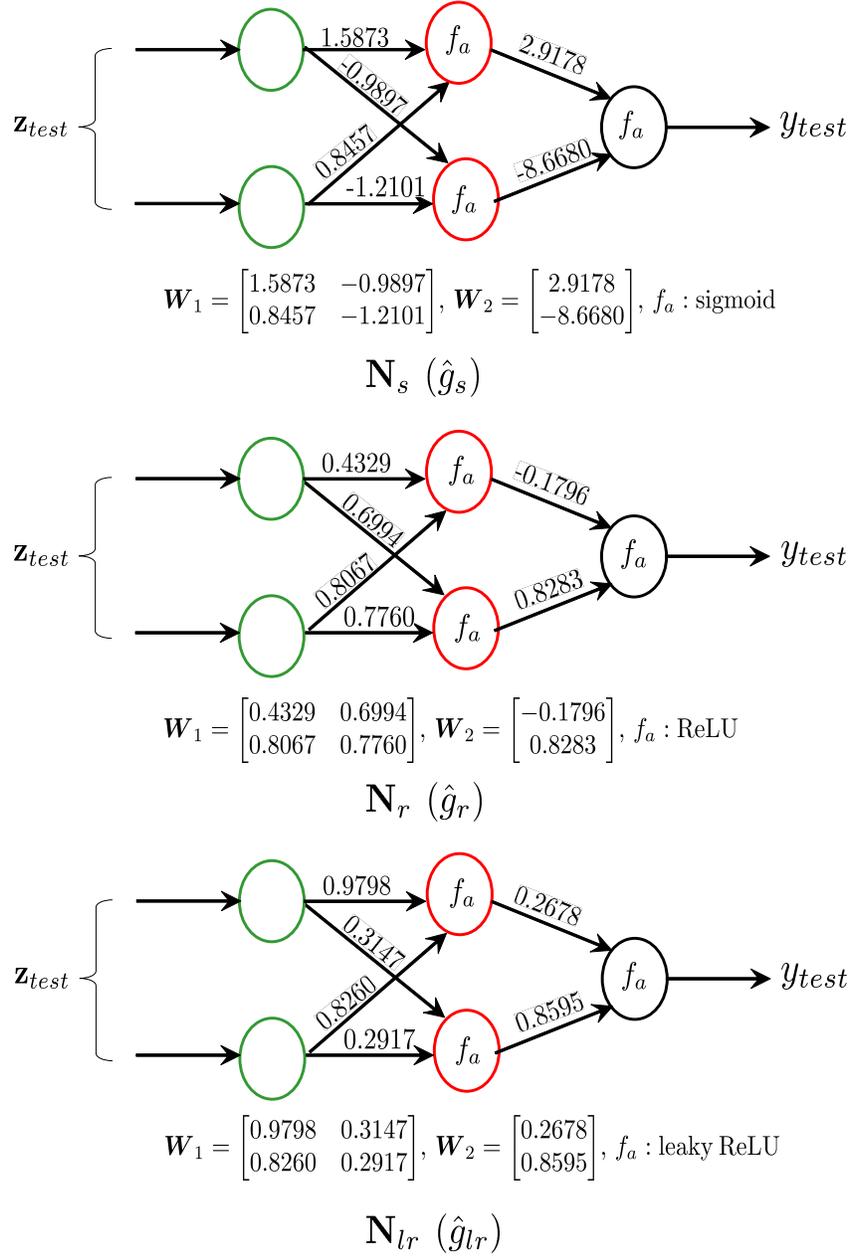}
\caption{The three trained DL networks ${\bf{N}}_s$, ${\bf{N}}_r$ and ${\bf{N}}_{lr}$ which generate the respective approximation functions ${\hat{g}}_s$, $\hat{g}_r$ and $\hat{g}_{lr}$, and by design ${\hat{g}}_s={\hat{g}}_r={\hat{g}}_{lr}=g$. Input, hidden and output layers are shown respectively in green, red and black color for clarity. Information about trained weights and activation functions is given below each network.}
\label{DLnetworks}
\end{figure}
In this lecture note, we used three commonly used activation functions namely sigmoid $f_a(x) = \frac{1}{1+e^{-x}}$, rectified linear unit (ReLU) $f_a(x)=max(0,x)$ and leaky ReLU $f_a(x)=\begin{cases}
    \alpha x,&x<0  \\
     x & x\geq0 \\
\end{cases}$. We thus obtained 3 trained DL networks namely ${\bf{N}}_s$, ${\bf{N}}_r$ and ${\bf{N}}_{lr}$ ($f_a=$ sigmoid, $f_a=$ ReLU and $f_a=$ leaky ReLU, respectively) with respective functional approximations ${\hat{g}}_s$, $\hat{g}_r$ and $\hat{g}_{lr}$. These networks have a single hidden layer ($L=1$), and are shown in Figure \ref{DLnetworks}. Another network namely ${\bf{N}}_{lr}^{(3)}$ (with approximation function $\hat{g}_{lr}^{(3)}$) was also trained with $L=3$ and $f_a=$ leaky ReLU. This is shown in Figure \ref{DL3}. Further, we employed $\ell=\text{MSE}$ for training all the 4 DL networks. Note that all these DL models/networks resulted in $\epsilon \approx 10^{-4}$ on the training dataset. Here, one also needs to ensure mitigation of the issue of overfitting (or memorization) i.e. a DL model performing well on training data but giving relatively high prediction error on a test set. To that end, a practical and well accepted solution is to examine the performance of the trained DL model on a test set which is independent of the training data \cite{10.5555/3086952}. Accordingly, we cross-validated the performance of trained DL model on an independent test dataset with 200 data points. We found that all the 4 trained DL models resulted in an error $\approx 10^{-4}$ i.e. the performance was similar to that on training dataset. Thus, ${\bf{N}}_s$, ${\bf{N}}_r$, ${\bf{N}}_{lr}$ and ${\bf{N}}_{lr}^{(3)}$ represent DL based models of $S$.

\subsubsection{A Closer Look at the Black box Problem}
As mentioned, explainable ML/DL seeks to get some insights into the black box (i.e. trained ML/DL model) by elaborating why a trained ML/DL model arrived at a particular prediction. However, there appears a lack of discussion on what really is meant by the black box nature of a trained DL model in the first place. Therefore, to understand the issue more closely, it is convenient to first analyze why $S$ as represented by \eqref{firlpfndomain} is not a black box.
Writing \eqref{firlpfndomain} in the frequency domain, we get:
\begin{equation}
Y\left(e^{j\Omega}\right) = 0.5\left(1+e^{-j\Omega}\right)X\left(e^{j\Omega}\right) = H\left(e^{j\Omega}\right)X\left(e^{j\Omega}\right)\,,
\label{dtft}
\end{equation}
where we use capital letters to denote the Discrete Time Fourier Transform (DTFT) of the corresponding signals. The symbol $\Omega$ represents the frequency of discrete signals (thus, principally $\Omega \in \left[-\pi,\pi\right]$).
\begin{figure}[] 
  \hspace{-4mm} 
  \subfloat[]{\includegraphics[scale=0.58]{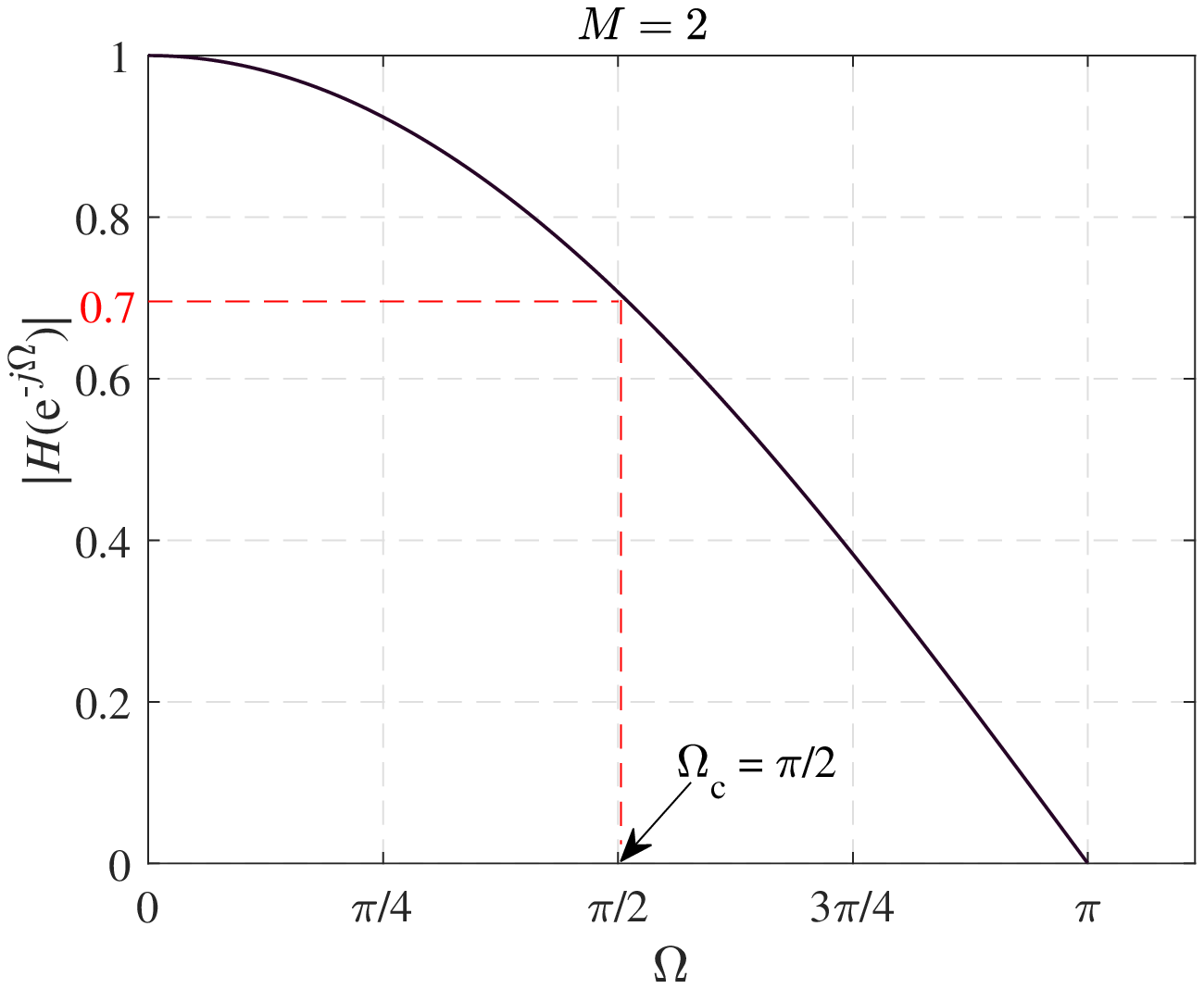} 
  \label{M2}
  } 
  \hspace{-4mm} 
  \subfloat[]{%
    \includegraphics[scale=0.59]{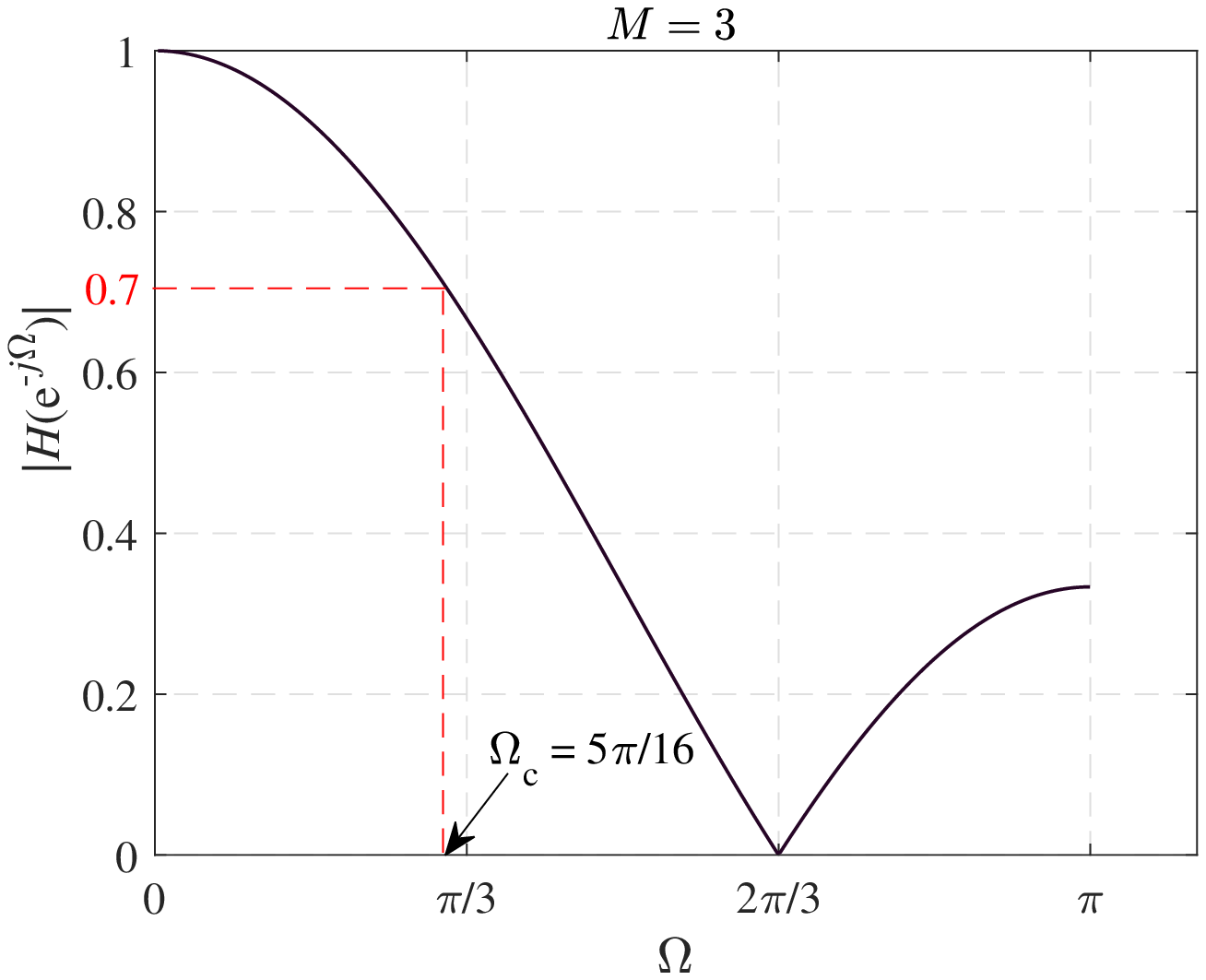} 
\label{M3}
  } 
  \caption{Plot of magnitude response in the range $\left[0,\pi\right]$ for two values of $M$. (a) $M=2$ with the cut-off frequency chosen as $\Omega_c=\frac{\pi}{2}$ (i.e. $2\:  \text{kHz}$), (b) $M=3$ with the cut-off frequency chosen as               $\Omega_c=\frac{5\pi}{16}$ (i.e. $1.25\:  \text{kHz}$). }
\label{dtftplot}
\end{figure}
Expression \eqref{dtft} tells us that the output $Y\left(e^{j\Omega}\right)$ of $S$ is simply an element wise multiplication of the input $X\left(e^{j\Omega}\right)$ with the system transfer function $H\left(e^{j\Omega}\right)$. Thus, observing the magnitude of $H\left(e^{j\Omega}\right)$ shown in Figure \ref{M2} allows us to understand the explicit nature of $S$. That is, the system $S$ essentially scales down (reduces) the strength of higher frequency components in the input. Such analysis is beneficial from a practical perspective because of the following reasons. First, it can provide prior meaningful insights and analysis about system output without actual implementation. For instance, a frequency $\Omega_0$ cannot occur in the output if it does not exist in the input. Second, it allows us to conceptualize and eventually design systems (algorithms) for say signal denoising or smoothing, anti-aliasing, equalizing and so on. This is enabled by the fact that undesired signals in the input (say noise or perceptually irrelevant components) can be appropriately characterized and then attenuated. As an example and as shown in Figure \ref{M2}, let us set the cut-off frequency $\Omega_c = \frac{\pi}{2}$ rad/sample i.e. $0.7\leq |H\left(e^{j\Omega}\right)|\leq 1$ for $0\leq \Omega \leq \frac{\pi}{2}$ (or equivalently $|H\left(e^{j\Omega}\right)|< 0.7$ for $\frac{\pi}{2}< \Omega \leq \pi$). Assuming the sampling frequency $f_s$ to be $8\:\text{kHz}$, we observe that $S$ attenuates, from a practical perspective, components beyond $2\: \text{kHz}$. It is, of course, possible to modify suitably the filtering characteristic depending on the application (eg. choosing another cut-off frequency). Thus, \eqref{dtft} and Figure \ref{M2} provide quantifiable insights into the working of $S$. Such insights also enable practically useful generalization of $S$ to other use-case scenarios. 

With the mentioned aspects of the system-centric approach in mind, it now becomes convenient to examine if the corresponding DL based model of $S$ is amenable to similar analysis or not. To that end, we take a closer look at the trained DL network ${\bf{N}}_r$ which is shown in Figure \ref{DLnetworks}, and begin by writing the explicit input output relationship. Let ${\bf{z}}_{test}=[x_1 \quad x_2]$ where $x_1,x_2 \in \mathbb{R}$. Then, using the weight matrices $W_1$ and $W_2$ of the trained model ${\bf{N}}_r$, we can write the expression for $y_{test}$ as:
\begin{equation}
y_{test} = max\Big\{0,\Big(0.8283\;max\left(0,0.6994\;x_1+0.7760\;x_2\right)-0.1796\;max\left(0,0.4329\;x_1+0.8067\;x_2\right)\Big)\Big\}\,,
\label{reludl}
\end{equation}



Now, by design, ${\bf{N}}_r$ is practically equivalent to $S$. This, in turn, implies that the corresponding function $\hat{g}_r$ defined by \eqref{reludl} essentially approximates a low pass filter by using a weighted combination of non linear $max(.)$ functions. However, establishing the low pass nature of ${\bf{N}}_r$ from analysis of \eqref{reludl} may be difficult, and this has serious implications from a practical perspective. Specifically, it means that we do not know how exactly the trained weights in ${\bf{N}}_r$ (refer to matrices $W_1$ and $W_2$) relate to its low pass filtering characteristic. Such lack of practically meaningful connection between the trained weights and their effect on the input signal essentially constitutes the black box nature of DL. One can similarly analyze the models ${\bf{N}}_s$, ${\bf{N}}_{lr}$ and ${\bf{N}}_{lr}^{(3)}$, and arrive at similar conclusions as that of ${\bf{N}}_r$.

\begin{figure}[]
\center
\includegraphics[scale=0.65]{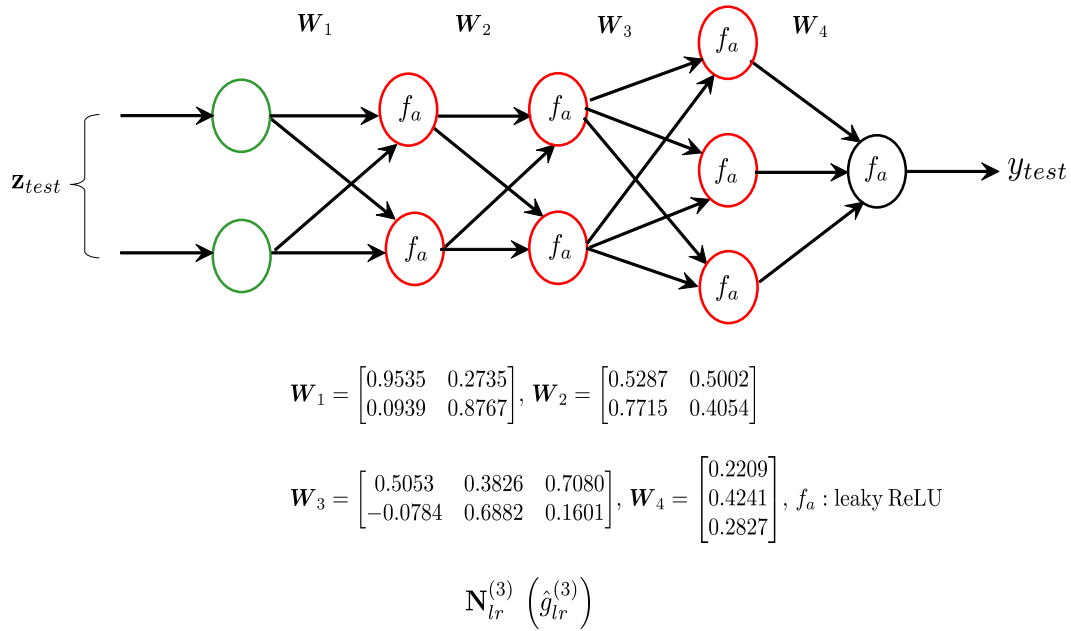}
\caption{Another trained DL network ${\bf{N}}_{lr}^{(3)}$ (with $L=3$) which generates the approximation function $\hat{g}_{lr}^{(3)}$ such that $\hat{g}_{lr}^{(3)} = g$. Input, hidden and output layers are shown respectively in green, red and black color. Information about the trained weights and activation function is also mentioned below the network. For visual clarity, the weights are not shown on the connections between neurons.}
\label{DL3}
\end{figure}

We will now analyze the limitations of the said black box issue from a practical perspective. Before doing that, we note that the presence of unknown and random noise in the training data $\left\{\left({\bf{z}}_i,{\bf{t}}_i\right)\right\}_{i=1,...,T}$ and the choice of the loss function (eg. MSE in our case) will also affect optimization. As a result, the weight matrices $\boldsymbol{W}_1$ to $\boldsymbol{W}_4$ (in Figures \ref{DLnetworks} and \ref{DL3}) might change. This will, however, not affect the analysis and conclusions made as long as the resultant DL models can approximate $S$.

\subsubsection{Practical implications of the black box problem}
Despite its seemingly theoretical underpinnings, a deeper understanding of the black box nature in ML/DL is also important from a practical view point. To elaborate on this, it is convenient to consider another use-case where perceptually irrelevant signal information now lies beyond say $1.25\:\: \text{kHz}$ (and not $2\:\: \text{kHz}$). Obviously, $S$ cannot directly serve the purpose in this case. However, from a system-centric perspective, it is fairly straightforward see that the required filter can be constructed by using $M=3$ in \eqref{firlpf}. Thereafter, an analysis similar to that of $S$ in \eqref{dtft} and Figure \ref{M2} can be carried out. Setting a cut-off frequency $\Omega_c = \frac{5\pi}{16}$ rad/sample, as visually illustrated in Figure \ref{M3}, will then result in the desired system $S{'}$. Thus, while $S$ and $S{'}$ are two different systems (filters) yet they are essentially unified via an interpretable and analytical philosophy. However, the situation is very different in case of ${\bf{N}}_r$. Specifically, there may not be a general and interpretable procedure of modifying the weights of ${\bf{N}}_r$ such that the resultant model say ${\bf{N}}'_r$ has a cut-off frequency of $1.25\: \text{kHz}$. The reason is, to reiterate, the lack of meaningful connection between the trained weights and the filtering characteristic of ${\bf{N}}_r$. Thus, the analysis of $S$ afforded by \eqref{dtft} and Figure \ref{M2} not only offers clear insights into $S$ but also provides a systematic approach to extend the scope to related practical use-cases. Likewise, other practical aspects such as controlling ripples in the passband or controlling gain in the transition band can be explicitly handled in $S$. By contrast, DL based modeling lacks such practically meaningful functionality due to the black box nature.  

Another notable practical limitation of the black box problem can show up when the ML/DL needs to be deployed in constrained environments where say latency, privacy or lack of communication bandwidth are important factors. In this use-case, the prediction and update of the DL trained model must take place with limited computing resources. For instance, on a local embedded processing near the sensor or on the edge servers  \cite{8763885}. For that reason, the number of parameters/weights of a DL model (which is one of the measures of model complexity\footnote{Model complexity also includes other components \cite{8763885} such as the number of mathematical operations needed, memory requirement, expressive capacity etc.}) must typically be reduced while maintaining practically reasonable prediction performance. In this context of on-device/on-edge computation, the black box nature can be a bottleneck in both DL model design and compression (reduction in number of weights). The reason is that the lack of clear and quantifiable connection between the trained weights and DL model functionality may prohibit a systematic understanding of the importance of the weights. As a result, making meaningful decisions about DL model reduction may become difficult and prone to trial and error. For instance, aspects such as which weights to quantize more (or less) and the extent of quantization, and/or which connections to prune (i.e. setting some of the weight values to 0) etc., may remain unclear. On the other hand, the system-centric approach can provide more flexibility toward explicitly analyzing the filtering characteristic of $S$ if $\boldsymbol{w}$ is changed due to any reason.     

The black box nature will also in general prohibit meaningful analysis of the DL model from the view point of its weaknesses (if any). Therefore, in this aspect, one might be constrained to merely analyze the output of DL and see if it matches the desired output or not (i.e. an almost exclusive focus on prediction accuracy). This potentially leaves room in terms of direct DL system analysis. In comparison, the system-centric approach allows a more direct system analysis. As an illustration, we may consider the system $S{'}$. As can be seen from Figure \ref{M3}, the transfer function of $S{'}$ shows a side lobe i.e. the magnitude increases slightly from $0$ at $\frac{2\pi}{3}$ to about $0.35$ at frequency $\pi$. Obviously, such side lobes should be as small as possible in the light of the use-case of attenuating information beyond $1.25\:\text{kHz}$. Hence, it represents a limitation of $S{'}$ and encourages steps for mitigating the same. A similar functionality in the corresponding DL based system say ${\bf{N}}_r$ might not be possible owing to the black box nature. 

Finally, the reader will note from Figure \ref{DLnetworks} that ${\bf{N}}_s$, ${\bf{N}}_r$ and ${\bf{N}}_{lr}$ have just 6 trainable weights. Hence, these DL models are not really \emph{complex} from the view point of dimensionality of the trained weights especially in comparison to several well known DL based architectures in vision (eg. a 50 layer Resnet \cite{7780459} has more than 25 million trained weights). Yet, all of them are essentially black boxes for the purpose at hand. Thus, the black box problem may not necessarily be attributed to high dimensionality of the trained weights alone. Instead, as we have illustrated, it is essentially a practical issue related to lack of meaningful association between the trained weights and the corresponding functionality of the DL model.   

\subsubsection{Why explainability is fundamental to DL system design?}

We note that the need for explainable ML/DL is largely fueled by high stakes applications \cite{xie2020explainable} like healthcare, autonomous driving, law enforcement, finance etc. Indeed, there is no denying that unexplained mistakes or wrong decisions made by a black box model in such applications can have serious implications. However, it is also equally interesting to note that explainability is in fact a fundamental concept for ML/DL system design, and not merely a post design requirement. In order to appreciate this aspect, it is once again convenient to think from a system-centric perspective, and consider our example of $S$. Observe that $S$ is uniquely characterized by \eqref{firlpfndomain} or equivalently via \eqref{dtft}. As a consequence, any analysis of $S$ (or even $S{'}$) from the perspective of its explainability, performance (strengths and/or weakness), implementation issues etc. can be carried out in an unambiguous fashion. However, this is not the case with DL based modeling of $S$ as all the four models, namely ${\bf{N}}_s$, ${\bf{N}}_r$, ${\bf{N}}_{lr}$ and ${\bf{N}}_{lr}^{(3)}$, accurately mimic the functionality of $S$. That is, we have $g={\hat{g}}_s={\hat{g}}_r={\hat{g}}_{lr}={\hat{g}}_{lr}^{(3)}$. In fact, one can train several other DL models by choosing different $L$, number of neurons in each layer, $f_a$ etc. Consequently, $g$ can be potentially approximated by a large number of DL models. This may seem useful at first sight. However, a closer scrutiny will reveal that such non unique DL based approximation leads to more questions than answers about the unknown function $g$. Particularly, one can raise the following questions:
\begin{enumerate}[\itshape(i)]
\item Which DL models' explanation should one rely upon to get accurate insights about the underlying system represented by $g$ and why?
\item Since all the DL models under question approximate $g$ quite well, should they all end up having same/similar explanations?  
\end{enumerate}
In the context of the first question, we may lack a systematic procedure to zero down on one DL model (out of several candidates such as ${\bf{N}}_s$, ${\bf{N}}_r$, ${\bf{N}}_{lr}$ and ${\bf{N}}_{lr}^{(3)}$) and its corresponding explanation as a surrogate to $g$. The second question also reveals interesting facet of DL modeling. If the answer to it is a yes, then again there may be difficulties in establishing the \emph{equivalence} of the said DL models which have different architectures (in terms of $L$, number of neurons in hidden layers and activation functions). On the other hand, if the answer to the second question above is a no, then we are potentially looking at a scenario where we have different DL based explanations for the same physical process (as defined by $g$ in our example). From a practical perspective, this may not be very meaningful. Thus, both the mentioned questions essentially emphasize why explainability should be an inherent and one of the first design principles in ML/DL system design, and not merely a \emph{posthoc} analysis procedure. 

%
%
%
\section{The case for system-centric philosophy based explainable ML}  
We have shown that the philosophy of explicit modeling can be leveraged to understand the black box nature and its practical implications in DL based systems. It is therefore logical to think that the same system-centric philosophy can also benefit DL based system design. Indeed, it is possible to exploit apriori domain knowledge and explicit tools both for DL system analysis and design. More specifically, from a DL system design perspective, there can be several practically useful implications toward:
\begin{enumerate}
\item preprocessing of the input data to overcome limitations of the DL model itself (eg. utilizing the idea of Fourier feature mapping to overcome the weakness of standard multilayer perceptron \cite{Tanciknips2020}).

\item incorporating aspects of an explicit system into learning based framework (eg. the use of \emph{deep unfolding} for image super resolution \cite{9157092} or in communication systems \cite{9186132}).

\item improving certain aspects of the DL model itself like performance (for instance, the use of Discrete Fourier Transform to improve pooling aspects \cite{Ryu_2018_ECCV} in Convolutional Neural Networks or understanding connections between DL and Graph Signal Processing \cite{9244181}).

\item avoiding training of the DL model from scratch and instead building upon simpler and interpretable aspects of known tools (eg. the Differentiable Digital Signal Processing (DDSP) framework \cite{engel2020ddsp} adapts interpretable DSP tools to diverse data via learning for audio applications).

\item visualizing the learning process in DL layers (eg. using Principal Component Analysis to detect adversarial examples \cite{8237877}), developing interactive visual analytics framework in the context of explainable DL (eg. \cite{8807299}, \cite{8402187}) or creating a design space of explainable systems for medical applications (eg. \cite{DBLP:conf/iui/XieCG19}).

\end{enumerate}
The above mentioned points (not an exhaustive list by any means) emphasize the potential benefits of incorporating a system-centric philosophy in the design, analysis and better performing but more explainable ML/DL systems. Essentially, such a strategy of desiging system-centric DL models can be a two-step process: a) developing a \emph{base} model which might rely more heavily on apriori domain knowledge, b) refinement of the resultant \emph{base} model via learning from application-specific data. This should lead to DL systems that may not only perform better (both in terms of generalization and accuracy) but are also transparent and amenable enough for a practically meaningful scrutiny.

\section{What we have learned} \label{conclusions}
The black box nature of ML/DL represents a fundamental problem. In this context, a system-centric perspective can lead to more in-depth understanding of this issue from the view point of its origin and practical implications. It also helps to appreciate why explainability represents a fundamental aspect of ML/DL system design. Such understanding may not only improve learning outcomes but can also provide meaningful insights toward improved practical deployment of ML/DL systems. Such deployment can span a broad canvas ranging from design considerations (including enhanced explainability, better generalization, proper model initialization and training etc.) to hardware implementations (on-device and under constrained environment).

\section{Acknowledgement}
The author acknowledges funding from the Science and Engineering Research Board (SERB), Department of Science and Technology, Government of India, vide grant no. SRG/2020/000849.

\bibliographystyle{IEEEtran}
\bibliography{publication}
%
%
%


\end{document}